\documentclass[12pt,a4paper,dvips]{article}
\usepackage{amsmath}
\usepackage{a4p,rotating}
\usepackage{epsfig,subfigure}
\usepackage{cite,mcite}
\usepackage{graphicx}
\usepackage{physics,l3_title,ifthen,Lep}
\journalname{Phys. Lett. B}
\date{February 6, 2002}
\preprint{2002-012}
\Lep{2}
\newlength{\capindent}
\newlength{\capwidth}
\newlength{\figwidth}
\newcommand{\icaption}[2][!*!,!]{\hspace*{\capindent}%
  \begin{minipage}{\capwidth}
    \ifthenelse{\equal{#1}{!*!,!}}%
      {\caption{#2}}%
      {\caption[#1]{#2}}
  \end{minipage}}
\newcommand{\pho}{\phantom{0}}

\newcommand{\EEEECCX}{\mathrm{ e^{+}e^{-} \rightarrow e^{+}e^{-}c\bar{c}X}}
\newcommand{\EEEEBBX}{\mathrm{ e^{+}e^{-} \rightarrow e^{+}e^{-}b\bar{b}X}}
\newcommand{\EEEEBBDX}{\mathrm{ e^{+}e^{-} \rightarrow e^{+}e^{-}b\bar{b}X \rightarrow D^{*\pm}X}}
\newcommand{\EEEEQQ}{\mathrm{ e^{+}e^{-} \rightarrow e^{+}e^{-}q\bar{q}}}
\newcommand{\EEEEDX}{\mathrm{ e^{+}e^{-} \rightarrow e^{+}e^{-}D^{*\pm}X}}

\newcommand{\GluonFusion}{\mathrm{ \gamma g \rightarrow c\bar{c}}}

\newcommand{\GG}{\mathrm{ \gamma \gamma}}
\newcommand{\EE}{\mathrm{ e^{+} e^{-}}}

\newcommand{\PT}{P_T}
\newcommand{\Ndata}{N_{obs}^{\Dst}}

\newcommand{\Dst}{\mathrm{ D^{*\pm}}}
\newcommand{\Dstpl}{\mathrm{ D^{*+}}}
\newcommand{\Dz}{\mathrm{ D^{0}}}

\newcommand{\DstToThreeProng}{\mathrm{ D^{*+} \rightarrow \Dz} \pi^+_s \rightarrow \mathrm{(K^{-} \pi^{+})}\pi^{+}_{s}}
\newcommand{\DstToPiZeroProng}{\mathrm{ D^{*+} \rightarrow \Dz} \pi^+_s \rightarrow \mathrm{(K^{-} \pi^{+} \pi^{0})}\pi^{+}_{s} }
\newcommand{\DstToFiveProng}{\mathrm{ D^{*+} \rightarrow \Dz} \pi^+_s
        \rightarrow \mathrm{(K^{-} \pi^{+} \pi^{-} \pi^{+})}\pi^{+}_{s} }
\newcommand{\PK}{\mathrm{(K^{-} \pi^{+})\pi^{+}_{s} }}
\newcommand{\PPPK}{\mathrm{(K^{-} \pi^{+} \pi^{-} \pi^{+})\pi^{+}_{s} }}
\newcommand{\PPzK}{\mathrm{(K^{-} \pi^{+} \pi^{0}) \pi^{+}_{s}}}
\newcommand{\DM}{\Delta M}

\newcommand{\VisEta}{\mathrm{\vert \eta \vert < 1.4}}
\newcommand{\VisPT}{1 \GeV < \PT < 12 \GeV}

\newcommand{\Annihilation}{\mathrm{  e^{+}e^{-} \rightarrow \gamma / Z \rightarrow q\bar{q}}}
\newcommand{\WPair}{\mathrm{  e^{+}e^{-} \rightarrow W^+ W^- \rightarrow q\bar{q}q\bar{q}}}

\newcommand{\dSigmadPT}{d\sigma / d\PT}
\newcommand{\dSigmadEta}{d\sigma / d\vert \eta \vert}
\newcommand{\dkpipiz}{\mathrm{D^0 \to K^- \pi^+ \pi^0 }}
\newcommand{\dzero}{\mathrm{ D^0}}

\def\PLB{{Phys. Lett.} {\bf B }}

\def\PRD{{Phys. Rev.} {\bf D }}
\def\ZPC{{Z. Phys.} {\bf C }}
\def\CPC{Comp. Phys. Comm. }

\begin{document}

\begin{titlepage}
\title{Inclusive $\mathrm{ \mathbf D^{*\pm}}$ Production    \\
        in Two-Photon Collisions at LEP}

\author{The L3 Collaboration}

%
%
\begin{abstract}
  
     Inclusive $\Dst$ production in two-photon collisions is studied with the 
L3 detector at LEP, using 683 $\mathrm{pb^{-1}}$
of data collected at centre-of-mass energies from 183 to $209 \GeV$. 
Differential cross sections are determined as functions 
of the transverse momentum and 
pseudorapidity of the $\Dst$ mesons in the kinematic 
region $\VisPT$ and $\VisEta$. The cross sections
 $\mathrm \sigma(\EEEEDX)$ 
in this kinematical region is measured and the $\mathrm \sigma(\EEEECCX)$
cross section is derived. The measurements are compared with 
next-to-leading order perturbative QCD calculations.
\end{abstract}
%
%
\submitted

\end{titlepage}
\normalsize

\section{Introduction}

\begin{figure}[ph]
\begin{center}
\includegraphics[width=1.\textwidth]{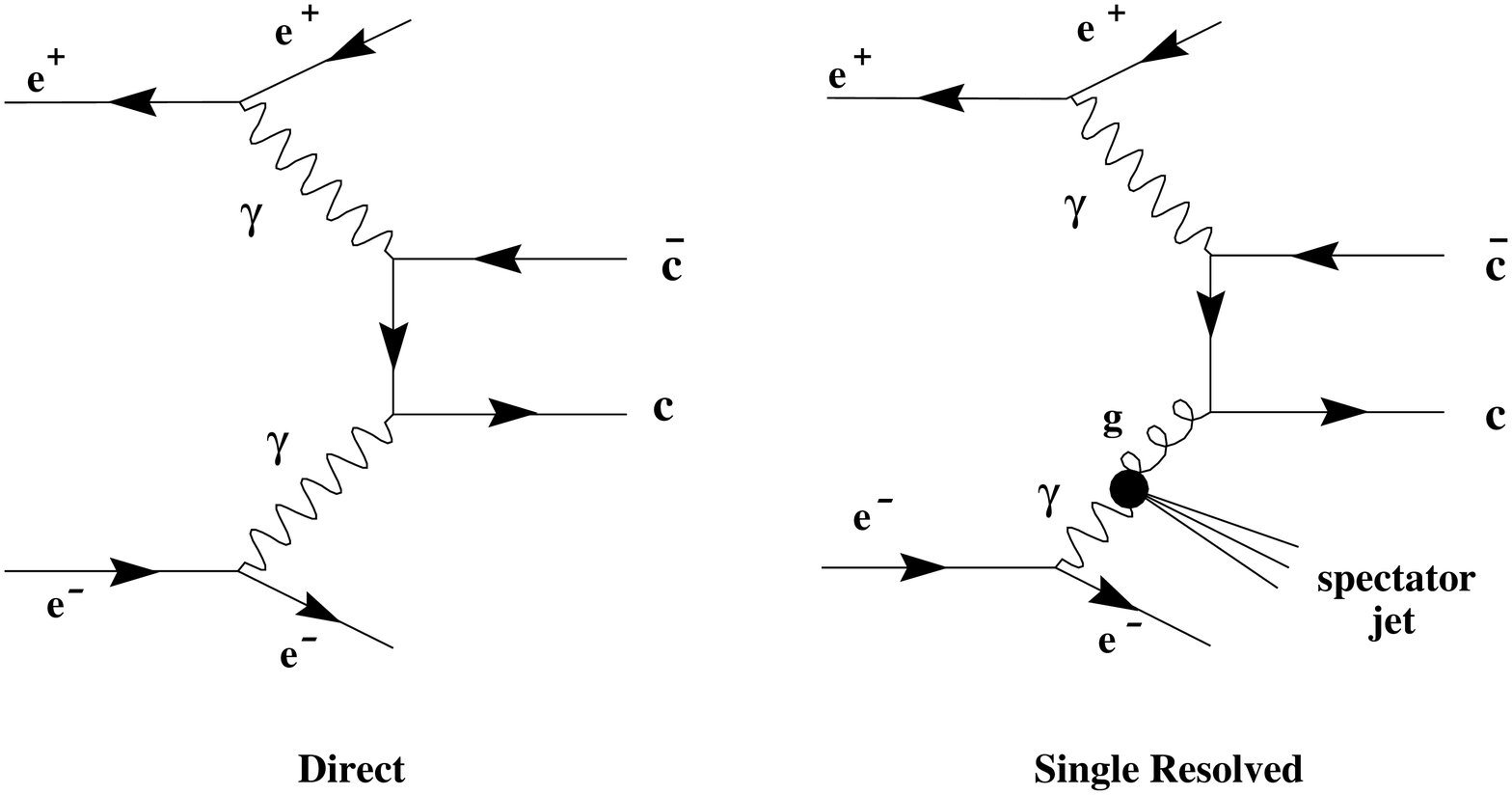}
\end{center}
\caption{Diagrams contributing to charm production in $\gamma\gamma$
collisions.}
\label{FeynDiagram}
\end{figure}

The measurement of open charm production in two-photon collisions 
provides a good test of perturbative QCD as the large physical scale 
set by the charm quark mass is expected to make the
perturbative calculations more reliable. At LEP2 energies, the direct and single resolved processes,
sketched in Figure \ref{FeynDiagram}, are predicted \cite{theory} to give 
comparable contributions to the cross section $\mathrm{\sigma(\GG \rightarrow c\bar{c})}$. 
The contributions to open charm production  
from soft processes described by the Vector Dominance Model (VDM) and from double resolved 
processes are expected 
to be small. The resolved photon cross section is dominated by the photon-gluon 
fusion process 
$\GluonFusion$. The cross section for open charm production in two-photon collisions therefore 
depends on the 
charm quark mass and on the gluonic parton density function of the photon. 
Measurements of open charm production 
in untagged two-photon collisions 
were performed at PEP\cite{TPC}, PETRA\cite{JADETASSO}, 
TRISTAN\cite{TOPVENAMY}, 
and LEP \cite{ALEPHL3,OPAL,L3,CCBB}, where 
charm quarks were identified by the presence of $\Dst$ mesons, 
leptons or $\mathrm{K^0_S}$ mesons.

This letter describes the results of a measurement of the $\EEEECCX$ cross 
section at 
centre-of-mass energies ranging from 183 to $209 \GeV$. 
The data sample corresponds 
to a total integrated luminosity 
of 683 $\mathrm{pb^{-1}}$ collected with the L3 detector\cite{L3detector} 
at LEP. 
Charm quarks are identified by reconstructing 
$\Dstpl$~\footnote{Charge conjugate states are assumed to be included.} 
mesons in three decay channels:
\begin{equation}
\DstToThreeProng
\label{ThreeProngMode}
\end{equation}
\begin{equation}
\DstToPiZeroProng
\label{PiZeroProngMode}
\end{equation}
\begin{equation}
\DstToFiveProng.
\label{FiveProngMode}
\end{equation}

\noindent $\pi_s^+$ 
stands for a ``slow pion'', so called as 
the kinetic energy release in $\Dstpl$ decays is only about $6 \MeV$.
Differential cross sections for $\Dst$ production as a function of its transverse momentum
$\PT$ and pseudorapidity $\eta$ are determined together with the visible cross section.
This is then extrapolated to derive $\mathrm \sigma(\EEEECCX)$.

The PYTHIA~\cite{PYTHIA} Monte Carlo program is used to model two-photon processes. 
It simulates $\GG$ events according 
to the current knowledge
of hadronic interactions obtained from pp and $\mathrm{\gamma p}$ studies. 
Two Monte Carlo samples are used: one 
with massless quark matrix elements, and another 
with matrix 
elements with a charm quark mass $m_c = 1.35 \GeV$. 
Both direct and resolved diagrams are included in the generation. The 
resolved processes are generated with the SaS1d photon parton density function~\cite{Schuler}. 
A two-photon luminosity function corresponding to the Equivalent 
Photon Approximation ~\cite{Budnev} is used  
with a cut off $\mathrm{Q^2 < m^2_\rho}$.
The L3 detector is simulated with the GEANT package~\cite{L3simulation} and 
Monte Carlo events are reconstructed in the same way as the data events. 
Time dependent detector inefficiencies, as 
monitored during the data taking period, are also simulated.

\section{Event Selection and $\mathrm{ \mathbf D^{*\pm}}$ Reconstruction}
The event selection follows two steps: first, hadronic final 
states from two-photon collisions are selected, 
then, $\Dst$ mesons are reconstructed.

\subsection{ Selection of Hadronic Two-Photon Events }
Hadronic events are required to contain at least four particles, 
with at least three charged tracks. A particle is defined as either 
a track in the central tracker 
or a calorimetric cluster not associated to a track. 
The background from $\mathrm{e^{+}e^{-} \ra e^{+}e^{-} \tau^{+} \tau^{-}}$ and
$\mathrm{e^{+}e^{-} \ra \tau^{+} \tau^{-}}$ events is highly suppressed by
this requirement.
To reject annihilation events, the visible energy has to be less than $80 \GeV$.
The analysis is restricted to events 
with visible mass $W_\mathrm{vis}>2 \GeV$,
as calculated from tracks and calorimetric clusters,
including those from the small angle luminosity monitor. 
All particles
are considered to be pions, except for unmatched electromagnetic clusters
considered as photons.

An anti-tagging condition is implemented on the scattered electron or positron,
by requiring the energy of the most energetic cluster in the forward 
electromagnetic calorimeter to be less than 0.2 $\mathrm{\sqrt{s}}$. 
Only 6$\%$ of events in the
$\Dst$ signal region are rejected by this criterium.
The anti-tagging requirement implies a small virtuality of the interacting 
photons: $\mathrm{\langle Q^2 \rangle \simeq 0.1} \GeV \mathrm{^2}$, as predicted by the PYTHIA Monte Carlo.
 
This preselection yields $4.6 \times 10^6$ events.
After applying the above cuts, the contamination from annihilation 
processes is less than 1$\%$. 

\subsection{ $\mathrm{ \mathbf D^{*\pm}}$ Reconstruction }

The $\Dst$ selection requires events with less than 15 tracks.
Tracks are required to have at least 20 hits, a transverse momentum
above $0.1 \GeV$ and a distance of closest approach to the interaction point 
in the transverse plane below 1.5 mm. 
In addition, photon candidates are selected in the electromagnetic calorimeter
as isolated showers with no close tracks.
$\Dst$ mesons are selected by first identifying $\Dz$ mesons. Additional criteria reject
the combinatorial background, and finally, $\pi_s^+$ candidates are added.
The $\Dz$  candidates are reconstructed from two-track combinations 
for the channel (\ref{ThreeProngMode}), 
two-track and two-photon combinations for the channel 
(\ref{PiZeroProngMode}), and four-track combinations for the channel (\ref{FiveProngMode}). 
Pion or kaon masses are assigned to the tracks. 
We define $P\mathrm{_K}$ and $P_\pi$ as the probabilities 
for the kaon and pion mass hypotheses, calculated from
the energy loss, $dE/dx$, in the tracker. The joint probabilities 
$P\mathrm{_K} \, P_\pi$ for channels 
(\ref{ThreeProngMode}) and (\ref{PiZeroProngMode}) and 
$P\mathrm{_K} \, P^1_\pi \, P^2_\pi \, P^3_\pi$ for channel (\ref{FiveProngMode}), 
have to be greater than 0.5$\%$ and 0.1$\%$ respectively.
In addition, for channel (\ref{FiveProngMode}) the kaon track 
should satisfy the condition $P\mathrm{_K} / P_\pi \ge 2$. 
The $\mathrm{\pi^0}$ candidates for channel (\ref{PiZeroProngMode}) 
are constructed from two neutral clusters 
in the barrel electromagnetic calorimeter,
with energy greater than $50 \MeV$
and effective mass within $\pm 20 \MeV$ of the nominal $\mathrm{\pi^0}$ mass. 
The transverse momentum of the $\mathrm{\pi^0}$ candidate with respect to the beam 
direction must be greater than $400 \MeV$.

The effective mass of the $\Dz$ candidate in the $\mathrm{K^{-} \pi^{+}}$,
$\mathrm{K^{-} \pi^{+} \pi^{0}}$, or $\mathrm{K^{-} \pi^{+} \pi^{-} \pi^{+} }$  combinations
is calculated 
for channels (\ref{ThreeProngMode}), (\ref{PiZeroProngMode}) and (\ref{FiveProngMode}), respectively. 
This mass should be within $\pm 75 \MeV$, $\pm 50 \MeV$, or $\pm 40 \MeV$ around the nominal $\Dz$ mass,
respectively. 
These different mass intervals are determined by the Monte Carlo $\Dz$ mass 
resolution 
and reflect the better transverse momentum resolution for softer tracks and 
the high precision $\mathrm{\pi^0}$ reconstruction in the BGO electromagnetic calorimeter. 

To reduce the combinatorial background in channel (\ref{ThreeProngMode}),
the opening angle $\mathrm{\alpha}$ between the kaon and the $\Dz$ flight direction has to satisfy the 
condition $\mathrm{\cos {\alpha} > -0.1}$.

A kinematical fit using as constraints the $\Dz$ and $\pi^0$ masses 
is performed for 
channel (\ref{PiZeroProngMode}). The fit
probability must be greater than 1$\%$. In addition, the sum of 
the transverse momenta squared of these particles must exceed $0.9 \GeV^2$.
Since the decay $\dkpipiz$ proceeds dominantly through one of the 
quasi-two-body intermediate states 
$\mathrm{\bar{K}^{*0} \pi^0}$,
$\mathrm{K^{*-} \pi^+}$ and 
$\mathrm{K^- \rho^+}$  \cite{PDG},
the $\mathrm{K^{-} \pi^+}$ and $\mathrm{K^{-} \pi^0}$ combinations
 are required to lie within $\pm 60 \MeV$ of the $\mathrm{K^*(892)}$ mass and the
$\mathrm{\pi^- \pi^0}$ ones within $\pm 150 \MeV$ of the $\mathrm{\rho}$ mass.
This last criterium also holds for channel (\ref{FiveProngMode}), that proceeds through the
$\mathrm{K \pi \rho}$ state.
In addition, for channel (\ref{PiZeroProngMode}) we require the helicity angle $\mathrm{\theta^*}$ of 
an intermediate decay state to satisfy the condition 
$\mathrm{\left| cos\theta^* \right| > 0.4}$, exploiting 
the P-wave properties of vector-particle decay into two scalar particles to reject background.
The helicity angle $\theta^*$ is defined as the angle between the direction of a decay product
of the vector resonance ($\mathrm{\bar{K}^{*0}}$, $\mathrm{K^{*-}}$ or $\rho^+$) and the 
direction of the pseudoscalar particle ($\pi^0$, $\pi^+$ or $\mathrm{K^-}$) from the $\dzero$ 
decay, calculated in the intermediate resonance rest frame.

$\dzero$ decay tracks are required to originate near the interaction point.
This is enforced by reconstructing a secondary vertex in the transverse plane  
for channels (\ref{ThreeProngMode}) and (\ref{FiveProngMode}).
The $\dzero$ impact parameter $d$ with respect to the interaction point in the transverse plane
must satisfy $d < 0.8$ mm and $d < 0.6$ mm respectively, while the $\dzero$
signed flight distance $s$ should satisfy $-0.8 < s < 1.0$ mm
and $-0.9 < s < 1.1$ mm, respectively.
No secondary vertex reconstruction is performed 
for channel (\ref{PiZeroProngMode}), but 
the distance of closest approach of kaon and pion tracks
to the interaction point in the transverse plane must be less than 1 mm.

Finally, a slow pion candidate track is added to the combination to form the $\Dst$ candidate.
The charge must be opposite to that of the kaon candidate,
with a distance of closest approach to the interaction point 
in the transverse plane less than 4.5 mm.

\begin{figure}[ph]
\begin{center}
\includegraphics[width=1.\textwidth]{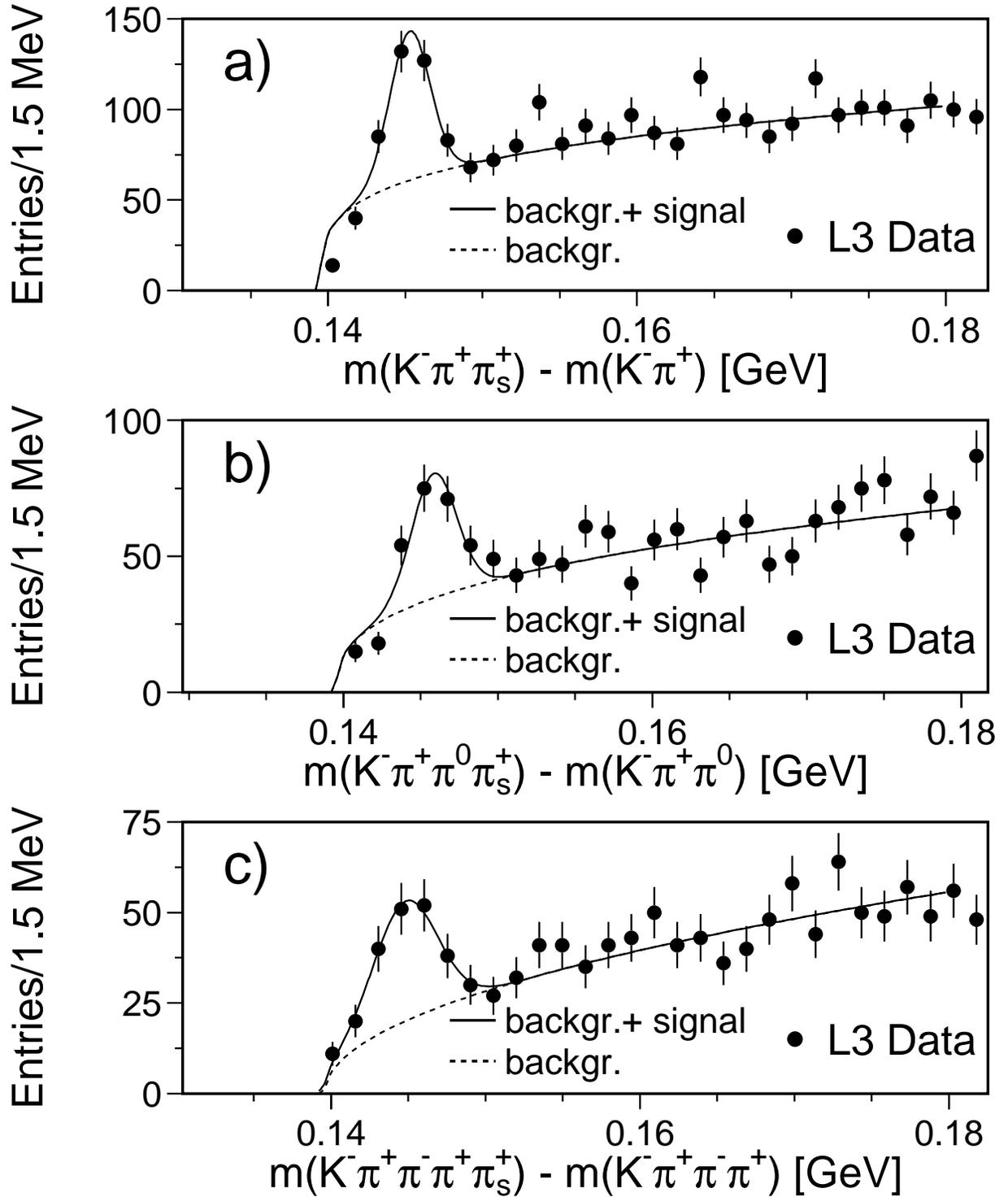}
\end{center}
\caption{Distribution of the mass difference $\Dstpl$ and $\Dz$
candidates for $\Dstpl$ decays into a) $\PK$, b) $\PPzK$ and c) $\PPPK$. 
The points are data, the error bars represent statistical 
uncertainties and the line is the result of the fit to the data
points.}
\label{SignalPicture}
\end{figure}

The mass difference $\DM$ of the $\mathrm{\Dst}$ and $\mathrm{\Dz}$ candidates 
is shown in Figure \ref{SignalPicture}.
$\Dst$ mesons are expected to form a peak around 
$145.4 \MeV$, while combinatorial background yields a rising distribution, starting at $139.6 \MeV$, 
the mass of the charged pion.
The reconstructed $\Dst$ signal is clearly observed in all the 
three decay channels under study. 
The number of observed $\Dst$ mesons is estimated by 
a fit to the $\DM$ spectrum. 
The background shape is modelled by Monte Carlo and 
checked using wrong-charge combinations in data, where the charge of the kaon candidate is the same
as the charge of the slow pion candidate. 
The background processes $\Annihilation$ and $\WPair$ represent less than 0.2$\%$ of the sample.
The background shape is modelled by
a function with two free parameters $a$ and $b$:
\begin{equation}
 f_{bkg}(\DM) = a ( \DM - m_\pi )^b,
\label{form.bkg}
\end{equation}
where the background shape parameter $b$ is determined from the $\EEEEQQ$ Monte Carlo.
A fit to the $\DM$ spectrum is performed, using 
a variable width Gaussian to describe the signal, and 
the fitted numbers of $\Dst$ mesons $\Ndata$ found in each channel are presented 
in Table \ref{XsectionsTable}.

\section{Results}

\subsection{ Differential cross sections }

The measured differential cross sections for inclusive $\Dst$ 
production as a function of its transverse momentum $\PT$ and
pseudorapidity $\eta$, $\dSigmadPT$ and $\dSigmadEta$, are 
shown in Tables \ref{tab.diff.pt} and \ref{tab.diff.eta}
for the visible region defined by:
\begin{equation}
1 \GeV{} < \PT  < 12 \GeV{}  \phantom{00} \mathrm{and} \phantom{00} \vert \eta \vert < 1.4 \phantom{.}. 
 \label{visible_region}
\end{equation}
When evaluating $\dSigmadPT$, which has a steep dependence on $\PT$, a 
correction factor, determined from Monte Carlo, is used to assign 
the differential cross sections to the centers of the corresponding bins.
The selection efficiency is determined with PYTHIA, using the $\Dstpl$
branching ratios of Reference~\citen{PDG} and  
assuming a $1 : 1$ mixture \cite{theory,OPAL} of direct and single resolved processes.
The sources of systematic uncertainty are discussed below and summarised in Table \ref{SystematicsTable}:
\begin{itemize}
        \item Selection procedure. The uncertainty is estimated by a variation of the selection criteria. 
        The largest contributions are the cut on the number of tracks~(6.0$\%$) 
and the $\mathrm{\Dz}$ mass window~(4.6$\%$) for channel~(\ref{ThreeProngMode}); 
the cut on the transverse momentum of the neutral pion~(7.8$\%$) and on the 
sum of transverse momenta~(4.3$\%$)
        for channel~(\ref{PiZeroProngMode});  
the cut on the $dE/dx$ probability ~(11.5$\%$) and the $\mathrm{\Dz}$ mass window~(7.5$\%$) 
for channel~(\ref{FiveProngMode}). 

        \item Branching ratios. The uncertainties are taken from Reference \citen{PDG}.

        \item Background. This uncertainty is 
        obtained by changing the background fit function within the uncertainties
        on its parameters. The range of the fit is also varied.

        \item Ratio of the direct to resolved processes.  
              The selection efficiency is re-evaluated changing the ratio 
              from $1 : 1$ to $3 : 1$ or $1 : 3$.

        \item Monte Carlo statistics.

        \item Trigger efficiency. The trigger efficiency and its uncertainty 
              are determined from the data using
              a set of independent triggers.

\end{itemize}

\begin{figure}[ph]
\begin{center}
\includegraphics[width=0.9\textwidth]{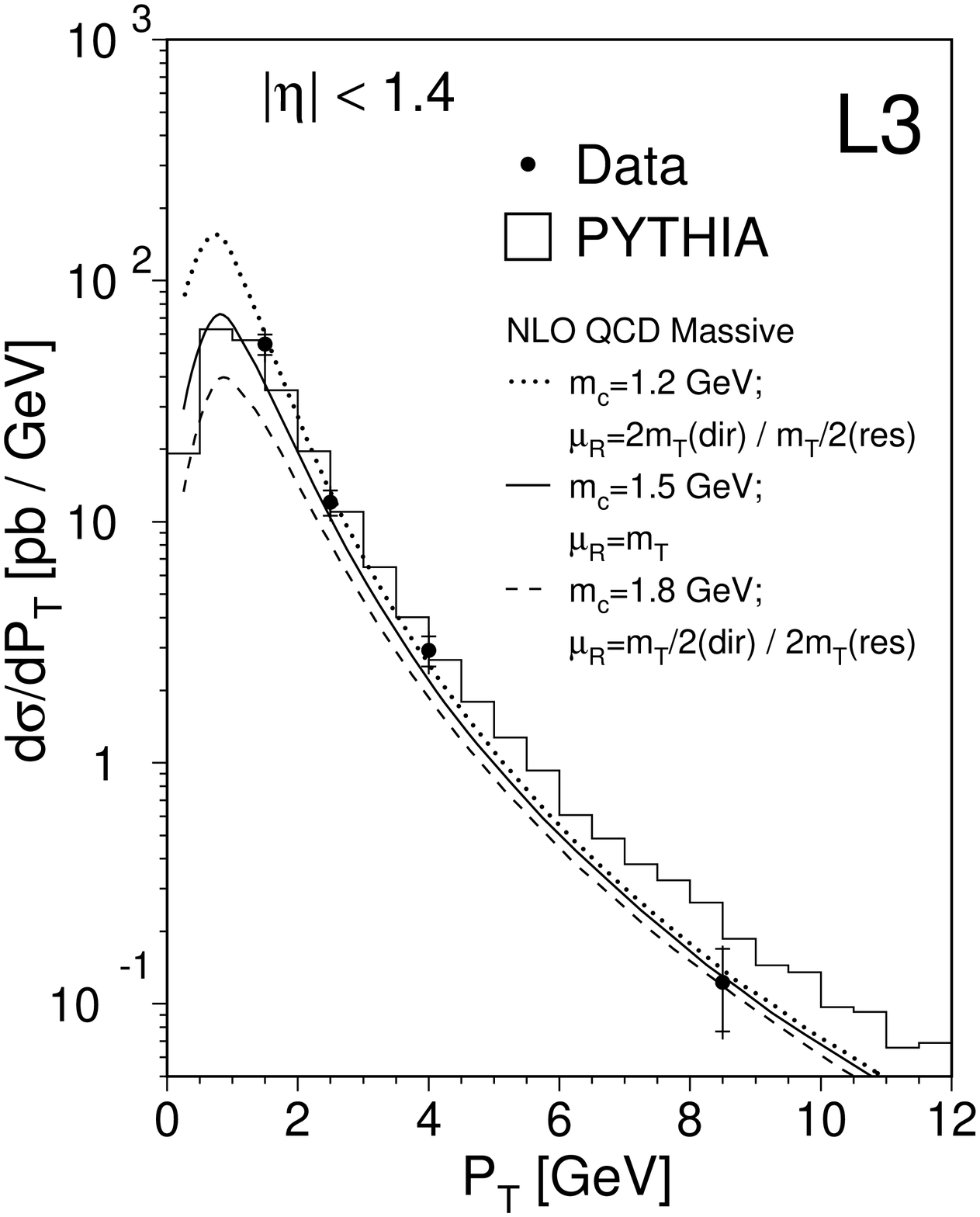}
\end{center}
\caption{ The differential cross section $\dSigmadPT$ for inclusive $\Dst$ production. 
The points are the data, the inner error 
bars represent the statistical uncertainty. 
The solid curve represents the NLO QCD calculations~\protect\cite{massQCD} 
with the Gl\"{u}ck-Reya-Schienbein~\protect\cite{GRS} parametrization 
of the parton density of the photon.
For the direct (dir) and resolved (res) processes, different 
values of the renormalisation 
and factorisation scales $ \mu_R$ and $ \mu_F$, as well 
as of $m_{\rm c}$, are used. 
The PYTHIA prediction is shown as a histogram.}
\label{fig.pt.diff}
\end{figure}

\begin{figure}[ph]
\begin{center}
\includegraphics[width=0.9\textwidth]{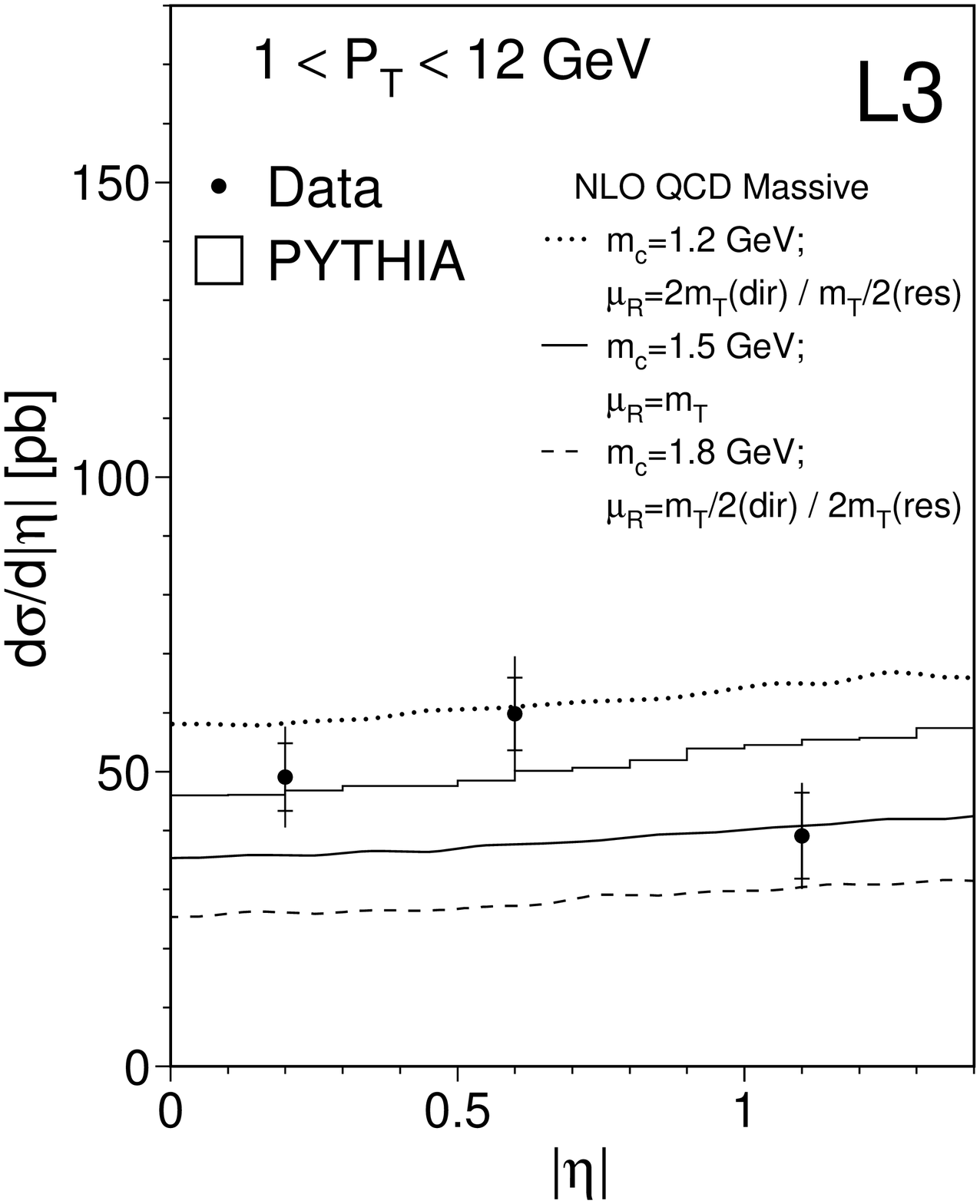}
\end{center}
\caption{ The differential cross section $\dSigmadEta$ for inclusive $\Dst$ production.
The points are the data, the inner error 
bars represent the statistical uncertainty. 
The solid curve represents the NLO QCD calculations~\protect\cite{massQCD} 
with the Gl\"{u}ck-Reya-Schienbein~\protect\cite{GRS} parametrization 
of the parton density of the photon.
For the direct (dir) and resolved (res) processes, different 
values of the renormalisation 
and factorisation scales $\mu_R$ and $ \mu_F$, as well 
as of $m_{\rm c}$, are used.
The PYTHIA prediction is shown as a histogram.} 
\label{fig.eta.diff}
\end{figure}

Figures \ref{fig.pt.diff} and \ref{fig.eta.diff} show the differential 
cross sections $\dSigmadPT$ and $\dSigmadEta$.
The differential cross sections
are compared to next-to-leading order (NLO) perturbative QCD calculations,
based on massive matrix elements \cite{massQCD}. 
In this scheme, the charm quark is not considered to be 
one of the active flavours inside the photon.
The Gl\"{u}ck-Reya-Schienbein  \cite{GRS} photon parton density 
is used
in the calculation.  
The renormalization scale, $ \mu_R$, and the factorization scale of the 
photon structure function, $ \mu_F$, are taken as 
$\mu_R =  \mu_F / 2 = m_T = \sqrt{ p_T^2 +m_{\rm{c}}^2}$,
where $ p_T$ is the transverse momentum of the charm quark whose mass is
$m_{\rm{c}} = 1.5 \GeV{}$. 
The calculations are repeated with different
renormalization scales separately for the direct and 
single-resolved contributions as well as with different charm quark masses, 
in order to estimate the maximal theory prediction uncertainty.
The measurements are in agreement with the NLO QCD calculations within 
the theoretical uncertainties.

\subsection{ Visible cross section }

The visible $\Dst$ production cross section $\mathrm{\sigma(\EEEEDX)}_{vis}$ is 
calculated in the kinematical region~(\ref{visible_region}) as:
\begin{equation}
  \sigma(\EEEEDX)_{vis} = \frac{\Ndata}{{\cal{L}} \varepsilon_{trig} \varepsilon^{\Dst}_{vis}},
\label{form.XSect}
\end{equation}  
where $\cal{L}$ is the total integrated luminosity. 
The trigger efficiency $\varepsilon_{trig}$ and 
the selection efficiency $\varepsilon^{\Dst}_{vis}$ 
are listed in Table \ref{XsectionsTable} and the number of observed events in 
Tables \ref{tab.diff.pt} and \ref{tab.diff.eta}.
Results for each $\PT$ and $\eta$ bin are shown in 
Tables \ref{tab.diff.pt} and \ref{tab.diff.eta},
while the integrated visible cross sections are given in Table \ref{XsectionsTable} for each channel.
Averaging over the three channels gives:
\begin{equation*}
        \sigma(\EEEEDX)_{vis} = 71.2 \pm 5.3 \pm 9.8 \; \mathrm{pb}, \notag
\end{equation*}
where the first uncertainty is statistical and the second is systematic.

\subsection{ Total cross section }

The total cross section of open charm production is calculated as:
\begin{equation}
        \mathrm{\sigma(\EEEECCX)} = \frac{\sigma(\EEEEDX)_{vis} - \sigma(\EEEEBBDX)_{vis}} 
{\varepsilon^{\Dst}_{tot}\, 2\, P(\mathrm{c \rightarrow D^{*+})}}.
\label{form.totxsec}
\end{equation}
The probability of a charm quark to hadronize into a $\Dstpl$ meson, 
$P(\mathrm{c \rightarrow D^{*+})}$, is 
equal to 0.241 $\pm$ 0.008 \cite{LEP.ElectroWeak}. The efficiency 
$\varepsilon^{\Dst}_{tot}$
is a correction factor to extrapolate from the visible $\Dst$ kinematical 
region~(\ref{visible_region}) to the full phase space. 
It follows from the calculations of Reference~\citen{massQCD} as 
$\varepsilon^{\Dst}_{tot} = 0.132 \; {^{+0.038}_{-0.043}}$.
The two-photon $\mathrm{b\bar{b}}$ production process also yields $\Dst$ mesons.
The cross section of this contribution, $\sigma(\EEEEBBDX)_{vis}$, is estimated 
from a $\EEEEBBX$ Monte Carlo sample and the measured
$\sigma(\EEEEBBX)$ cross section \cite{CCBB} as $1.9 \pm 0.4$ pb.
It is subtracted and 
its uncertainty is included in the systematics.

The total open charm production cross section at the average 
centre-of-mass energy of $197\GeV$ is estimated to be:
\begin{equation*}
\mathrm{
        \sigma(\EEEECCX) = (1.12 \pm 0.09 \pm 0.16 \; {^{+0.54}_{-0.25}} \;) \times 10^3 \; pb.
} \notag
\end{equation*}
The first uncertainty is statistical and second is systematic.
The third uncertainty is that on the extrapolation 
from the visible phase space region to the full one.
It corresponds 
to the integration~\cite{massQCD} of the dotted and
dashed lines in Figures \ref{fig.pt.diff} and \ref{fig.eta.diff}.
A consistent estimate of this uncertainty is also obtained by 
comparing the extrapolation factors obtained using different
Monte Carlo generators: 
the massive and massless PYTHIA samples yield extrapolation factors 
that differ by 24$\%$.

\begin{figure}[ph]
\begin{center}
\includegraphics[width=1.\textwidth]{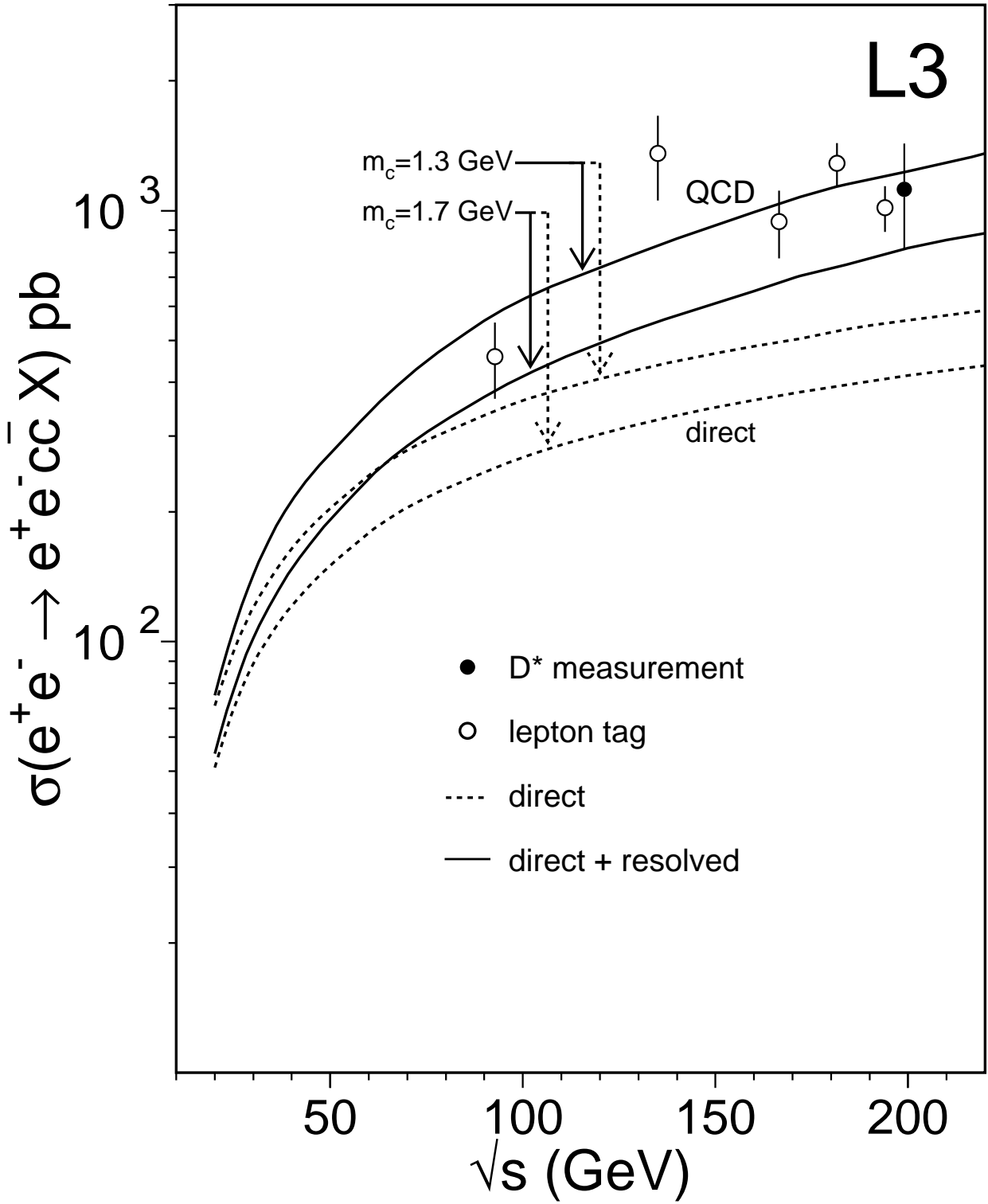}
\end{center}
\caption{The open charm production cross section in two-photon collisions
as a function of the $\EE$ centre-of-mass energy. 
L3 measurements based on lepton tag \protect\cite{L3,CCBB} are plotted as
the open circles.
The dashed lines correspond to the direct process contribution 
and the solid lines represent the NLO QCD prediction \protect\cite{theory}
for the sum of the direct and resolved processes. 
The effect of different values of $m_{\rm c}$
is also shown.}
\label{TotalXSectPicture}
\end{figure}

The total inclusive charm production cross section as a function of 
the $\EE$ centre-of-mass energy is shown in Figure \ref{TotalXSectPicture} 
together with our previous measurements \cite{L3,CCBB}. 
The data are compared to the NLO QCD predictions of 
Reference \citen{theory} for the direct process  
and the sum of direct and resolved processes.
In this calculation, the charm quark mass is taken to be
$1.3 \GeV$ or $1.7 \GeV$, and the open charm threshold to be at $3.8 \GeV$.
The resolved process is calculated using the Gl\"{u}ck-Reya-Vogt~\cite{GRV}
photon parton density function. The renormalization and factorization 
scales are taken equal to the charm quark mass.
The use of the Drees-Grassie \cite{DG} parton density function 
results in a decrease of the cross
section of 9\% for $m_{\rm c} = 1.3 \GeV$ and of 3\% 
for $m_{\rm c} = 1.7 \GeV$.
Changing the QCD scale from $m_{\rm c}$ to 2$m_{\rm c}$ 
decreases the predicted cross section by 30\% for $m_{\rm c} = 1.3 \GeV$ 
and 15\% for $m_{\rm c} = 1.7 \GeV$.

The measured cross section is in good agreement with 
our previous measurements ~\cite{L3,CCBB}
based on lepton tag and with the QCD 
expectations. The contribution of the
single-resolved process is needed to describe the data.

\section*{Acknowledgements }
We thank S. Frixione, E. Laenen and M. Kr\"{a}mer 
for providing us with the results of QCD cross section calculations.

\clearpage
  
\bibliographystyle{l3stylem}

\clearpage
\newpage
%
%
\section*{Author List}
\typeout{   }     
\typeout{Using author list for paper 251 -- ? }
\typeout{$Modified: Jul 15 2001 by smele $}
\typeout{!!!!  This should only be used with document option a4p!!!!}
\typeout{   }
%
%
%
%
%
%

\newcount\tutecount  \tutecount=0
\def\tutenum#1{\global\advance\tutecount by 1 \xdef#1{\the\tutecount}}
\def\tute#1{$^{#1}$}
\tutenum\aachen            
\tutenum\nikhef            
\tutenum\mich              
\tutenum\lapp              
\tutenum\basel             
\tutenum\lsu               
\tutenum\beijing           
\tutenum\berlin            
\tutenum\bologna           
\tutenum\tata              
\tutenum\ne                
\tutenum\bucharest         
\tutenum\budapest          
\tutenum\mit               
\tutenum\panjab            
\tutenum\debrecen          
\tutenum\florence          
\tutenum\cern              
\tutenum\wl                
\tutenum\geneva            
\tutenum\hefei             
\tutenum\lausanne          
\tutenum\lyon              
\tutenum\madrid            
\tutenum\florida           
\tutenum\milan             
\tutenum\moscow            
\tutenum\naples            
\tutenum\cyprus            
\tutenum\nymegen           
\tutenum\caltech           
\tutenum\perugia           
\tutenum\peters            
\tutenum\cmu               
\tutenum\potenza           
\tutenum\prince            
\tutenum\riverside         
\tutenum\rome              
\tutenum\salerno           
\tutenum\ucsd              
\tutenum\sofia             
\tutenum\korea             
\tutenum\purdue            
\tutenum\psinst            
\tutenum\zeuthen           
\tutenum\eth               
\tutenum\hamburg           
\tutenum\taiwan            
\tutenum\tsinghua          

{
\parskip=0pt
\noindent
{\bf The L3 Collaboration:}
\ifx\selectfont\undefined
 \baselineskip=10.8pt
 \baselineskip\baselinestretch\baselineskip
 \normalbaselineskip\baselineskip
 \ixpt
\else
 \fontsize{9}{10.8pt}\selectfont
\fi
\medskip
\tolerance=10000
\hbadness=5000
\raggedright
\hsize=162truemm\hoffset=0mm
\def\r{\rlap,}
\noindent

P.Achard\r\tute\geneva\ 
O.Adriani\r\tute{\florence}\ 
M.Aguilar-Benitez\r\tute\madrid\ 
J.Alcaraz\r\tute{\madrid,\cern}\ 
G.Alemanni\r\tute\lausanne\
J.Allaby\r\tute\cern\
A.Aloisio\r\tute\naples\ 
M.G.Alviggi\r\tute\naples\
H.Anderhub\r\tute\eth\ 
V.P.Andreev\r\tute{\lsu,\peters}\
% ...
\newpage
\begin{list}{A}{\itemsep=0pt plus 0pt minus 0pt\parsep=0pt plus 0pt minus 0pt
                \topsep=0pt plus 0pt minus 0pt}
\item[\aachen]
 I. Physikalisches Institut, RWTH, D-52056 Aachen, FRG$^{\S}$\\
 III. Physikalisches Institut, RWTH, D-52056 Aachen, FRG$^{\S}$
\item[\nikhef] National Institute for High Energy Physics, NIKHEF, 
     and University of Amsterdam, NL-1009 DB Amsterdam, The Netherlands
\item[\mich] University of Michigan, Ann Arbor, MI 48109, USA
\item[\lapp] Laboratoire d'Annecy-le-Vieux de Physique des Particules, 
     LAPP,IN2P3-CNRS, BP 110, F-74941 Annecy-le-Vieux CEDEX, France
\item[\basel] Institute of Physics, University of Basel, CH-4056 Basel,
     Switzerland
\item[\lsu] Louisiana State University, Baton Rouge, LA 70803, USA
\item[\beijing] Institute of High Energy Physics, IHEP, 
  100039 Beijing, China$^{\triangle}$ 
\item[\berlin] Humboldt University, D-10099 Berlin, FRG$^{\S}$
\item[\bologna] University of Bologna and INFN-Sezione di Bologna, 
     I-40126 Bologna, Italy
\item[\tata] Tata Institute of Fundamental Research, Mumbai (Bombay) 400 005, India
\item[\ne] Northeastern University, Boston, MA 02115, USA
\item[\bucharest] Institute of Atomic Physics and University of Bucharest,
     R-76900 Bucharest, Romania
\item[\budapest] Central Research Institute for Physics of the 
     Hungarian Academy of Sciences, H-1525 Budapest 114, Hungary$^{\ddag}$
\item[\mit] Massachusetts Institute of Technology, Cambridge, MA 02139, USA
\item[\panjab] Panjab University, Chandigarh 160 014, India.
\item[\debrecen] KLTE-ATOMKI, H-4010 Debrecen, Hungary$^\P$
\item[\florence] INFN Sezione di Firenze and University of Florence, 
     I-50125 Florence, Italy
\item[\cern] European Laboratory for Particle Physics, CERN, 
     CH-1211 Geneva 23, Switzerland
\item[\wl] World Laboratory, FBLJA  Project, CH-1211 Geneva 23, Switzerland
\item[\geneva] University of Geneva, CH-1211 Geneva 4, Switzerland
\item[\hefei] Chinese University of Science and Technology, USTC,
      Hefei, Anhui 230 029, China$^{\triangle}$
\item[\lausanne] University of Lausanne, CH-1015 Lausanne, Switzerland
\item[\lyon] Institut de Physique Nucl\'eaire de Lyon, 
     IN2P3-CNRS,Universit\'e Claude Bernard, 
     F-69622 Villeurbanne, France
\item[\madrid] Centro de Investigaciones Energ{\'e}ticas, 
     Medioambientales y Tecnol\'ogicas, CIEMAT, E-28040 Madrid,
     Spain${\flat}$ 
\item[\florida] Florida Institute of Technology, Melbourne, FL 32901, USA
\item[\milan] INFN-Sezione di Milano, I-20133 Milan, Italy
\item[\moscow] Institute of Theoretical and Experimental Physics, ITEP, 
     Moscow, Russia
\item[\naples] INFN-Sezione di Napoli and University of Naples, 
     I-80125 Naples, Italy
\item[\cyprus] Department of Physics, University of Cyprus,
     Nicosia, Cyprus
\item[\nymegen] University of Nijmegen and NIKHEF, 
     NL-6525 ED Nijmegen, The Netherlands
\item[\caltech] California Institute of Technology, Pasadena, CA 91125, USA
\item[\perugia] INFN-Sezione di Perugia and Universit\`a Degli 
     Studi di Perugia, I-06100 Perugia, Italy   
\item[\peters] Nuclear Physics Institute, St. Petersburg, Russia
\item[\cmu] Carnegie Mellon University, Pittsburgh, PA 15213, USA
\item[\potenza] INFN-Sezione di Napoli and University of Potenza, 
     I-85100 Potenza, Italy
\item[\prince] Princeton University, Princeton, NJ 08544, USA
\item[\riverside] University of Californa, Riverside, CA 92521, USA
\item[\rome] INFN-Sezione di Roma and University of Rome, ``La Sapienza",
     I-00185 Rome, Italy
\item[\salerno] University and INFN, Salerno, I-84100 Salerno, Italy
\item[\ucsd] University of California, San Diego, CA 92093, USA
\item[\sofia] Bulgarian Academy of Sciences, Central Lab.~of 
     Mechatronics and Instrumentation, BU-1113 Sofia, Bulgaria
\item[\korea]  The Center for High Energy Physics, 
     Kyungpook National University, 702-701 Taegu, Republic of Korea
\item[\purdue] Purdue University, West Lafayette, IN 47907, USA
\item[\psinst] Paul Scherrer Institut, PSI, CH-5232 Villigen, Switzerland
\item[\zeuthen] DESY, D-15738 Zeuthen, 
     FRG
\item[\eth] Eidgen\"ossische Technische Hochschule, ETH Z\"urich,
     CH-8093 Z\"urich, Switzerland
\item[\hamburg] University of Hamburg, D-22761 Hamburg, FRG
\item[\taiwan] National Central University, Chung-Li, Taiwan, China
\item[\tsinghua] Department of Physics, National Tsing Hua University,
      Taiwan, China
\item[\S]  Supported by the German Bundesministerium 
        f\"ur Bildung, Wissenschaft, Forschung und Technologie
\item[\ddag] Supported by the Hungarian OTKA fund under contract
numbers T019181, F023259 and T024011.
\item[\P] Also supported by the Hungarian OTKA fund under contract
  number T026178.
\item[$\flat$] Supported also by the Comisi\'on Interministerial de Ciencia y 
        Tecnolog{\'\i}a.
\item[$\sharp$] Also supported by CONICET and Universidad Nacional de La Plata,
        CC 67, 1900 La Plata, Argentina.
\item[$\triangle$] Supported by the National Natural Science
  Foundation of China.
\end{list}
}
\vfill


\clearpage

\begin{table}
\begin{center}
\small
\begin{tabular}{|l|l|l|l|}
\hline
Channel                      & $\PK$                   & $\PPzK$                & $\PPPK$                \\
\hline
$\Ndata$                   & 245\phantom{.} $\pm$ 27  & 139\phantom{.0} $\pm$ 18  & 99\phantom{.0} $\pm$ 16 \\
$\varepsilon_{trig}$, $\%$ & 92.9 $\pm$ 0.2          & 92.1 \pho $\pm$ 0.3         & 89.4 $\pm$ 0.2         \\
$\varepsilon^{\Dst}_{vis}$, $\%$ 
                             &   0.61 $\pm$  0.03      &  0.22 \pho $\pm$  0.02      &  0.20 $\pm$  0.02      \\
\hline
 $\sigma^{\Dst}_{vis}$, pb& 63.3 $\pm$ 6.7 \pho $\pm$ 8.9 & 100.9  $\pm$ 13.0 $\pm$ 18.2  & 81.1 $\pm$ 12.5 $\pm$ 17.8  \\
$\mathrm \sigma(\EEEECCX)$, nb& 1.00 $\pm$ 0.11 $\pm$ 0.15${^{+0.48}_{-0.22}}$
                                      & 1.59 \pho $\pm$ 0.21 $\pm$ 0.31${^{+0.77}_{-0.36}}$
                                               & 1.28 $\pm$ 0.20 $\pm$ 0.29${^{+0.62}_{-0.29}}$ \\
\hline
\end{tabular}
\caption{ The number of observed $\Dst$ mesons for the three different channels and 
the trigger ($\varepsilon_{trig}$) and
visible ($\varepsilon^{\Dst}_{vis}$) efficiencies.
Visible and total cross sections
are also given. 
The uncertainties on the numbers of reconstructed $\Dst$ mesons and 
on the efficiencies are statistical. 
For the cross section values, the first uncertainty 
is statistical, the second is systematic. 
The last uncertainty for the total charm cross section
corresponds to the extrapolation to the total phase space.} 
\label{XsectionsTable}
\end{center}
\normalsize
\end{table}

\small
\begin{table}
\begin{center}
\begin{tabular}{|c|c|c|c|}
\hline
$\PT$$\mathrm{(GeV)}$   & $\Ndata$ & $\dSigmadPT$ ($\mathrm{pb/GeV}$) & $\mathrm{\Delta\sigma}$ (pb) \\
\hline
1 $-$ \pho 2 & 325 $\pm$ 31      &  $54.5\pho     \pm  5.3\pho \pm  6.7\pho$             & 54.5   $\pm$ 5.3   $\pm$ 6.7  \\
2 $-$ \pho 3 & 130 $\pm$ 17      &  $12.1\pho     \pm  1.4\pho \pm  1.4\pho$            & 12.1   $\pm$ 1.4   $\pm$ 1.4  \\
3 $-$ \pho 5 & \pho 72 $\pm$ 11  &  $\pho 2.9\pho \pm  0.4\pho \pm 0.4\pho$          & \pho 6.7 $\pm$ 1.0  $\pm$ 1.0 \\
5 $-$ 12  & \pho 13 $\pm$ \pho 5 &  $\pho 0.12    \pm 0.05 \pm 0.02$                  & \pho 1.5 $\pm$ 0.6 $\pm$ 0.3 \\
\hline
\end{tabular}
\end{center}
  \caption{Number of observed $\Dst$ mesons $\Ndata$ and 
measured $\Dst$ production cross sections in different $\PT$ bins
for $\VisEta$. Differential $\dSigmadPT$ cross sections
refer to the center of the bin and $\mathrm{\Delta\sigma}$ is the integral over the bin. }
  \label{tab.diff.pt} 
\end{table}

\begin{table}
\begin{center}
\begin{tabular}{|c|c|c|c|}
\hline
$\mathrm{\left|\eta\right|}$  & $\Ndata$ & $\mathrm{d\sigma / d\vert \eta \vert}$ (pb) & $\mathrm{\Delta\sigma}$ (pb)  \\
\hline
0.0 $-$ 0.4   & 194 $\pm$ 23  & 49.4 $\pm$ 6.0 $\pm$ 6.0 & 19.8 $\pm$ 2.4 $\pm$ 2.4  \\
0.4 $-$ 0.8        & 232 $\pm$ 24  & 60.3 $\pm$ 6.0 $\pm$ 7.2 & 24.1 $\pm$ 2.4 $\pm$ 2.9  \\
0.8 $-$ 1.4        & \pho 92 $\pm$ 17  & 39.4 $\pm$ 7.2 $\pm$ 5.6 & 23.6 $\pm$ 4.3 $\pm$ 3.4  \\
\hline
\end{tabular}
\end{center}
  \caption{Number of observed $\Dst$ mesons $\Ndata$ and 
measured $\Dst$ production cross sections
for $\VisPT$. Differential $\mathrm{d\sigma / d\vert \eta \vert}$ cross sections
refer to the center of the bin and $\mathrm{\Delta\sigma}$ is the integral over the bin. }
  \label{tab.diff.eta} 
\end{table}

\begin{table}
\begin{center}
  \begin{tabular}{|l|c|c|c|}
\hline
Source              &  $\PK$      &  $\PPzK$     &  $\PPPK$         \\
\hline
Selection           & \pho 7.8     &  \pho 9.9      &  16.8       \\
Branching Ratio~\protect\cite{PDG} & \pho 3.1   & \pho 7.2    & \pho 4.9        \\
Fit procedure   & \pho 6.6     &  \pho 6.6      & \pho 6.6         \\ 
Direct/Resolved ratio & \pho 8.6   &  \pho 8.6     & \pho 8.6        \\
Monte Carlo statistics   & \pho 5.1    &  \pho 9.4     & \pho 8.7          \\
Trigger efficiency  & \pho 0.2     &  \pho 0.3      & \pho 0.2            \\
\hline
Systematics visible   & 14.6    &  18.9    &  22.3          \\
\hline
$P(\mathrm{c \rightarrow D^{*+})}$
                    & \pho 3.1    & \pho 3.1    & \pho 3.1         \\
$\mathrm{b\bar{b}}$
 contribution       & \pho 0.6     & \pho 0.6     & \pho 0.6          \\

\hline
Systematics total   & 14.9    &  19.2    &  22.5          \\
\hline
\end{tabular}
\caption{Sources of systematic uncertainty in \%
on the visible $\Dst$ cross section and the total charm cross section. 
The total systematic uncertainty is calculated by 
adding all contributions in quadrature.}
\label{SystematicsTable}
\end{center}
\end{table}

\clearpage
\vfil
\vfil

\clearpage
\vfil
\vfil

\clearpage
\vfil
\vfil

\clearpage
\vfil
\vfil

\clearpage
\vfil
\vfil

\end{document}